\documentclass[twocolumn]{aastex631}

\usepackage{amsmath}
\usepackage{amssymb}
\usepackage{bm}
\usepackage{physics}
\usepackage{booktabs}
\usepackage{multirow}
\usepackage{xspace}
\usepackage{graphicx}

\shorttitle{Stellar Halo Memory}
\shortauthors{Pandey, and Mondal}

\def\refjnl#1{{\rm#1}}
\def\apjl{\refjnl{ApJ Letters}}                
\def\apjs{\refjnl{ApJS}}               
\def\ao{\refjnl{Appl.~Opt.}}           
\def\aap{\refjnl{A\&A}}                

\begin{document}

\title{Stellar Halo Memory: A New Observable for Galactic Archaeology}


\author{Biswajit Pandey}
\affiliation{Department of Physics, Visva-Bharati University, Santiniketan, 731235, India}

\author{Amit Mondal}
\affiliation{Department of Physics, Visva-Bharati University, Santiniketan, 731235, India}


\correspondingauthor{Biswajit Pandey}
\email{biswap@visva-bharati.ac.in}


\begin{abstract}

The Galactic stellar halo preserves a fossil record of the Milky Way's assembly history, yet there is currently no unified framework for quantifying how much of this information survives in the present-day Galaxy. We introduce an information-theoretic framework that establishes stellar halo memory as a directly measurable observable for Galactic archaeology. Using APOGEE Data Release 17 halo stars, we quantify the statistically significant information shared among stellar dynamical and chemical observables through the excess mutual information, enabling a physically motivated decomposition into dynamical, chemical, and cross-memory reservoirs. We construct cumulative radial memory profiles together with two dimensionless diagnostics, the memory dominance ratio and the coupling efficiency, to investigate how the surviving assembly memory is partitioned throughout the Galactic halo. We find that the different memory observables exhibit distinct radial evolution in the inner halo but all evolve toward a common statistically significant residual memory state beyond $\sim20$\,kpc. Comparison with randomized realizations demonstrates that the measured memories exceed the corresponding null expectation by large factors, confirming that they represent genuine astrophysical structure rather than statistical fluctuations. The surviving assembly memory is consistently dominated by dynamical correlations while retaining a finite chemo-dynamical coupling across the halo. These results suggest that phase mixing redistributes, rather than completely erases, the memory imprinted during galaxy formation. Our study establishes stellar halo memory as a new observable for Galactic archaeology and provides a unified statistical framework for investigating galaxy assembly in both observations and cosmological simulations.

\end{abstract}

\section{Introduction}

The stellar halo of the Milky Way is the longest-lived record of the Galaxy's formation. Within the hierarchical paradigm of $\Lambda$CDM cosmology, it is assembled through the continuous accretion of satellite galaxies together with a smaller \textit{in situ} stellar component, preserving the cumulative imprint of billions of years of dynamical evolution and chemical enrichment \citep{bullock05,cooper10,helmi20}. Every accreted progenitor contributes stars with characteristic orbital properties and elemental abundance patterns that reflect their birth environments. As these systems are disrupted, tidal stripping, phase mixing, and violent relaxation progressively erase their dynamical coherence, while stellar chemical abundances remain largely unchanged after formation. Consequently, the present-day stellar halo retains only a partial memory of its assembly, encoded jointly in stellar dynamics, chemical abundances, and the statistical correspondence between them. A fundamental objective of Galactic archaeology is therefore not merely to identify the relics of past accretion events, but to determine how much of this primordial memory survives and how it is distributed throughout the stellar halo.

The past two decades have witnessed remarkable progress toward reconstructing the assembly history of the Milky Way. Precision astrometry from {\it Gaia}, together with large spectroscopic surveys such as APOGEE, GALAH, LAMOST, and Gaia-ESO, has transformed Galactic archaeology into a precision discipline capable of probing the fossil record of galaxy formation \citep{gaia18,gaia23,majewski17}. Dynamical studies have uncovered stellar streams, shells, and coherent structures that preserve the signatures of past accretion events \citep{johnston98,helmi99,belokurov06,helmi20}. At the same time, chemical abundance analyses have established elemental abundance patterns as powerful tracers of stellar birth environments and nucleosynthetic histories, motivating the development of chemical tagging and chemo-dynamical approaches to identify stars of common origin \citep{freeman02,ting15}. Together, these complementary approaches have revealed an increasingly detailed picture of the Galaxy's formation history.

Despite these advances, the surviving assembly memory of the stellar halo has never been treated as a single physical entity. Dynamical studies quantify orbital coherence, chemical studies characterize abundance distributions, and chemo-dynamical investigations examine correlations between the two, yet these observables are almost always analysed independently using different statistical methodologies. This separation obscures a more fundamental view of galaxy assembly. Every galaxy formation event simultaneously imprints three distinct, but intrinsically coupled, signatures on the stellar halo. The first is imprinted in phase space through the orbital configuration of stars and evolves under gravitational dynamics. The second is imprinted in chemical abundance space through the nucleosynthetic history of the progenitor system and is largely preserved over stellar lifetimes. The third resides in the statistical correspondence between dynamics and chemistry, encoding how stellar birth environments map onto present-day orbital structure. Accordingly, we identify three complementary memory reservoirs: \emph{dynamical memory}, \emph{chemical memory}, and \emph{cross memory}. Together, they constitute the surviving memory of galaxy assembly. Quantifying their relative importance and mutual coupling remains an outstanding challenge in Galactic archaeology.

Information theory \citep{shannon48} provides a natural language for addressing this problem. Shannon mutual information measures the amount of information shared between two random variables and provides a model-independent measure of statistical dependence that is sensitive to arbitrary nonlinear relationships. Unlike conventional correlation statistics, it requires no assumption regarding the functional form of the underlying dependence and therefore provides a natural description of the complex, multidimensional distributions encountered in stellar halo populations. From this perspective, galaxy formation may be regarded as an information-generating process that establishes statistical dependencies among stellar dynamical and chemical properties, while subsequent evolution redistributes and progressively erases these dependencies through phase mixing and relaxation. The mutual information preserved by the present-day stellar halo therefore provides a direct statistical measure of its surviving assembly memory.

Although the application of information theory to Galactic archaeology remains largely unexplored, information-theoretic methods have previously been employed in cosmology to investigate statistical memory in the large-scale structure of the Universe and to quantify the dependence between galaxy properties and their environments \citep{pandey17,sarkar20}. Motivated by these developments, we extend the information-theoretic perspective to Galactic archaeology by introducing \emph{stellar halo memory} as a directly measurable observable that quantifies how information about galaxy assembly is preserved within the chemo-dynamical structure of the Milky Way.

In this Letter, we develop and apply a unified information-theoretic framework for measuring stellar halo memory using APOGEE DR17 observations \citep{abdurrouf22}. We define three complementary classes of memory corresponding to the dynamical, chemical, and cross-memory reservoirs described above. Using halo stars from APOGEE Data Release~17, we estimate cumulative radial profiles of the \emph{excess mutual information}, defined as the observed mutual information after subtracting the expectation obtained from randomized realizations to remove finite-sample bias. We further introduce two derived diagnostics: the \emph{memory dominance ratio}, which quantifies the relative importance of dynamical and chemical memory, and the \emph{coupling efficiency}, which measures the strength of their mutual association. Together, these quantities establish a unified observational framework for measuring the surviving memory of the Galactic stellar halo. More broadly, they recast Galactic archaeology as the quantitative study of how information generated during galaxy formation is preserved, redistributed, and ultimately remembered by the stellar halo.

\section{Stellar Halo Memory}

The principal objective of Galactic archaeology is to infer the formation history of the Milky Way from the statistical properties of its present-day stellar populations. This requires observables that retain measurable signatures of past evolutionary processes despite billions of years of dynamical evolution. We propose that the \emph{memory} preserved within the stellar halo constitutes such an observable. Unlike individual stellar streams or chemically homogeneous stellar populations, stellar halo memory is a global statistical property of the halo that quantifies how much information about galaxy assembly remains encoded in the joint distribution of stellar dynamical and chemical observables.

We define the stellar halo memory associated with two physical observables $X$ and $Y$ as the statistically significant shared information that remains encoded between them in the present-day stellar halo. Operationally, this observable is estimated from the excess mutual information,

\begin{equation}
\mathcal{M}(X,Y)
\equiv
I_{\rm exc}(X;Y)
=
I(X;Y)-I_{\rm null}(X;Y),
\label{eq:memory}
\end{equation}

where $I(X;Y)$ is the observed mutual information and $I_{\rm null}(X;Y)$ is the expectation obtained after randomly permuting one of the variables. The subtraction removes the finite-sample bias of the estimator and isolates the statistically significant information that cannot be attributed to chance associations. Throughout this Letter, stellar halo memory is the physical observable of interest, while excess mutual information provides its estimator.

Galaxy assembly simultaneously imprints three distinct yet complementary forms of stellar halo memory. We therefore decompose the observable into three physically motivated memory reservoirs. The \emph{dynamical memory},

\begin{equation}
\mathcal{M}_{\rm D}
=
\left\{
\mathcal{M}(r,v_r),
~
\mathcal{M}(E,L_z)
\right\},
\end{equation}

quantifies the residual phase-space coherence preserved by gravitational evolution, where $r$, $v_r$, $E$, and $L$ denote the Galactocentric radius, radial velocity, orbital energy, and angular momentum, respectively. The \emph{chemical memory},

\begin{equation}
\begin{split}
\mathcal{M}_{\rm C}
&=
\left\{
\mathcal{M}([\mathrm{Fe/H}], [\alpha/\mathrm{Fe}]), \mathcal{M}([\mathrm{Fe/H}], [\mathrm{Mg/H}]), 
\right. \\
&\quad \left. \mathcal{M}([\mathrm{Fe/H}], [\mathrm{O/H}]))
\right\},
\end{split}
\end{equation}

measures the statistical coherence of stellar chemical abundances that preserve the nucleosynthetic histories of their progenitor systems. Finally, the \emph{cross memory},

\begin{equation}
\mathcal{M}_{\rm X}
=
\mathcal{M}(Y,Z),
\end{equation}

where $Y$ and $Z$ denote a dynamical variable and a chemical abundance tracer respectively, measures the surviving chemo-dynamical memory. It therefore quantifies the extent to which present-day orbital structure continues to retain information about stellar birth environments. 

To investigate how the memory of the stellar halo varies with Galactocentric distance, we evaluate the observable cumulatively using all stars interior to radius $R$, yielding the radial memory profile,

\begin{equation}
\mathcal{M}(<R),
\end{equation}

which measures the total assembly memory enclosed within radius $R$. This cumulative construction suppresses statistical fluctuations while directly tracing the radial build-up of stellar halo memory.

The observable is estimated using the non-parametric $k$-nearest-neighbour mutual information estimator of \citet{kraskov04}, which accurately captures nonlinear statistical dependencies without requiring density estimation through binning. Prior to estimation, every variable is standardized independently within each cumulative sample. Throughout this work we adopt $k=4$, although our conclusions are insensitive to reasonable variations of this choice. The null memory is obtained by repeatedly permuting one variable while preserving its marginal distribution, and the ensemble-averaged mutual information defines $I_{\rm null}$ in \autoref{eq:memory}.

To characterize how the stellar halo partitions its surviving memory, we introduce two dimensionless diagnostics. The \emph{memory dominance ratio},

\begin{equation}
\mathcal{R}
=
\frac{\mathcal{M}_{\rm D}}
{\mathcal{M}_{\rm C}},
\end{equation}

quantifies whether the retained memory is predominantly dynamical ($\mathcal{R}>1$) or chemical ($\mathcal{R}<1$). We further define the \emph{coupling efficiency},

\begin{equation}
\eta
=
\frac{\mathcal{M}_{\rm X}}
{\mathcal{M}_{\rm D}
+
\mathcal{M}_{\rm C}},
\label{eq:efficiency}
\end{equation}

which measures the fraction of the total stellar halo memory that remains encoded in the coupling between dynamics and chemistry. Together, $\mathcal{R}$ and $\eta$ provide a compact physical description of how the memory of galaxy assembly is partitioned among the three fundamental memory reservoirs.

The remainder of this Letter is therefore concerned with the radial behaviour of stellar halo memory and its implications for the assembly history of the Milky Way.

\section{Data}

Measuring stellar halo memory requires accurate determinations of both stellar chemical abundances and phase-space coordinates. We therefore combine high-resolution spectroscopy from APOGEE with precision astrometry from \textit{Gaia} to construct a chemo-dynamical sample of Galactic halo stars.

Our analysis is based on the Apache Point Observatory Galactic Evolution Experiment (APOGEE) Data Release 17 \citep{majewski17,abdurrouf22}, which provides near-infrared spectra ($R\simeq22\,500$) together with homogeneous stellar atmospheric parameters and elemental abundances derived using the APOGEE Stellar Parameters and Chemical Abundance Pipeline (ASPCAP; \citealt{perez16}). Following the quality recommendations of \citet{horta23}, we select stars satisfying $\mathrm{S/N}>70$, $3500<T_{\rm eff}<6500\,\mathrm{K}$, \texttt{STARFLAG}=0, and $\texttt{ASPCAPFLAG}<256$. We adopt the broader surface gravity range $0<\log g<5$ in order to retain both giant and dwarf halo stars. For each star we use the spectroscopic quantities \texttt{FE\_H}, \texttt{ALPHA\_M}, \texttt{M\_H}, \texttt{MG\_FE}, \texttt{O\_FE}, \texttt{SI\_FE}, and the heliocentric radial velocity \texttt{VHELIO\_AVG}.

The APOGEE catalogue is cross-matched with \textit{Gaia} EDR3 \citep{gaia21} to obtain celestial coordinates, parallaxes, and proper motions. Distances are adopted from the photogeometric estimates of \citet{bailer21} (\texttt{GAIAEDR3\_R\_MED\_PHOTOGEO}), which combine \textit{Gaia} astrometry with photometric information and provide reliable distance estimates over the large range of heliocentric distances probed by the halo population. We further apply the magnitude-dependent proper-motion corrections of \citet{cantat21} before deriving stellar kinematics.

Halo stars are identified using combined kinematic and chemical criteria. Following \citet{gaia18}, we compute the transverse velocity,

\begin{equation}
V_T=\frac{4.7405}{\varpi}
\sqrt{pmra^2+pmdec^2},
\end{equation}

where $\varpi$ is the parallax (mas), and $pmra$ and $pmdec$ are the proper motions in right ascension and declination, respectively. We retain stars with $V_T>200~{\rm km\,s^{-1}}$, which efficiently suppresses contamination from the Galactic disc while preserving a predominantly halo population \citep{gaia18}. To further reduce contamination from metal-rich disc stars, we additionally require ${\rm [Fe/H]}<-1$ \citep{conroy19}. The resulting sample provides the six-dimensional phase-space information together with homogeneous chemical abundances required to quantify the dynamical, chemical, and cross memory of the Galactic stellar halo.

\section{Results}

\begin{figure*}
\centering
\includegraphics[width=0.8\textwidth]{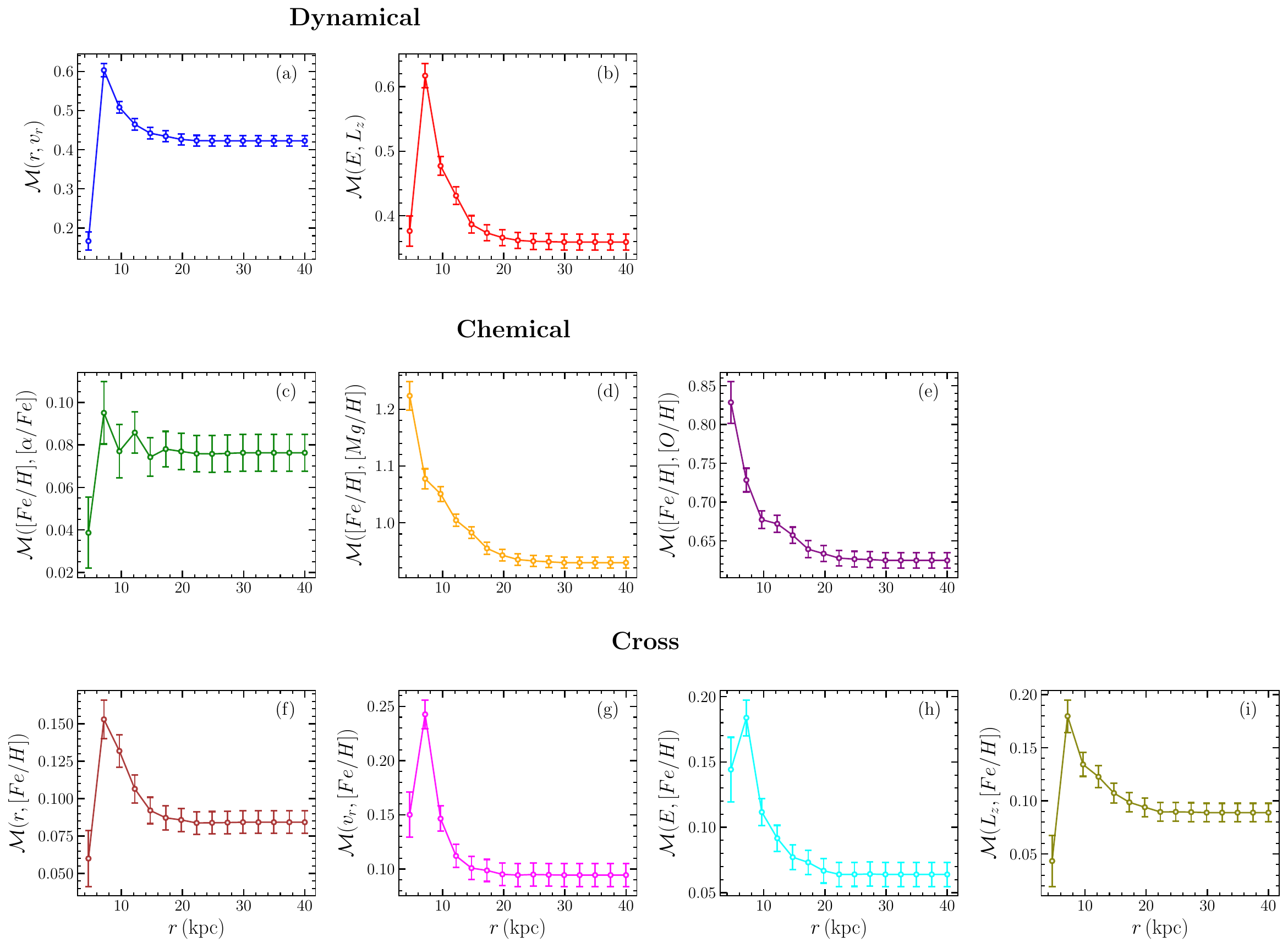}
\caption{Radial profiles of the cumulative stellar halo memory measured from APOGEE DR17 halo stars. The observables are organized into the three memory reservoirs introduced in Section~2: \emph{Dynamical Memory} (panels a--b), \emph{Chemical Memory} (panels c--e), and \emph{Cross Memory} (panels f--i). Each profile is computed cumulatively using all stars interior to the Galactocentric radius $r$, with error bars representing $1\sigma$ uncertainties derived via jackknife resampling. Although the individual memory observables exhibit diverse radial evolution, all converge toward nearly constant values beyond $r\sim20$\,kpc, indicating the emergence of a statistically significant residual memory state in the outer stellar halo.}
\label{fig:memory}
\end{figure*}

\begin{figure*}
\centering
\includegraphics[width=0.8\textwidth]{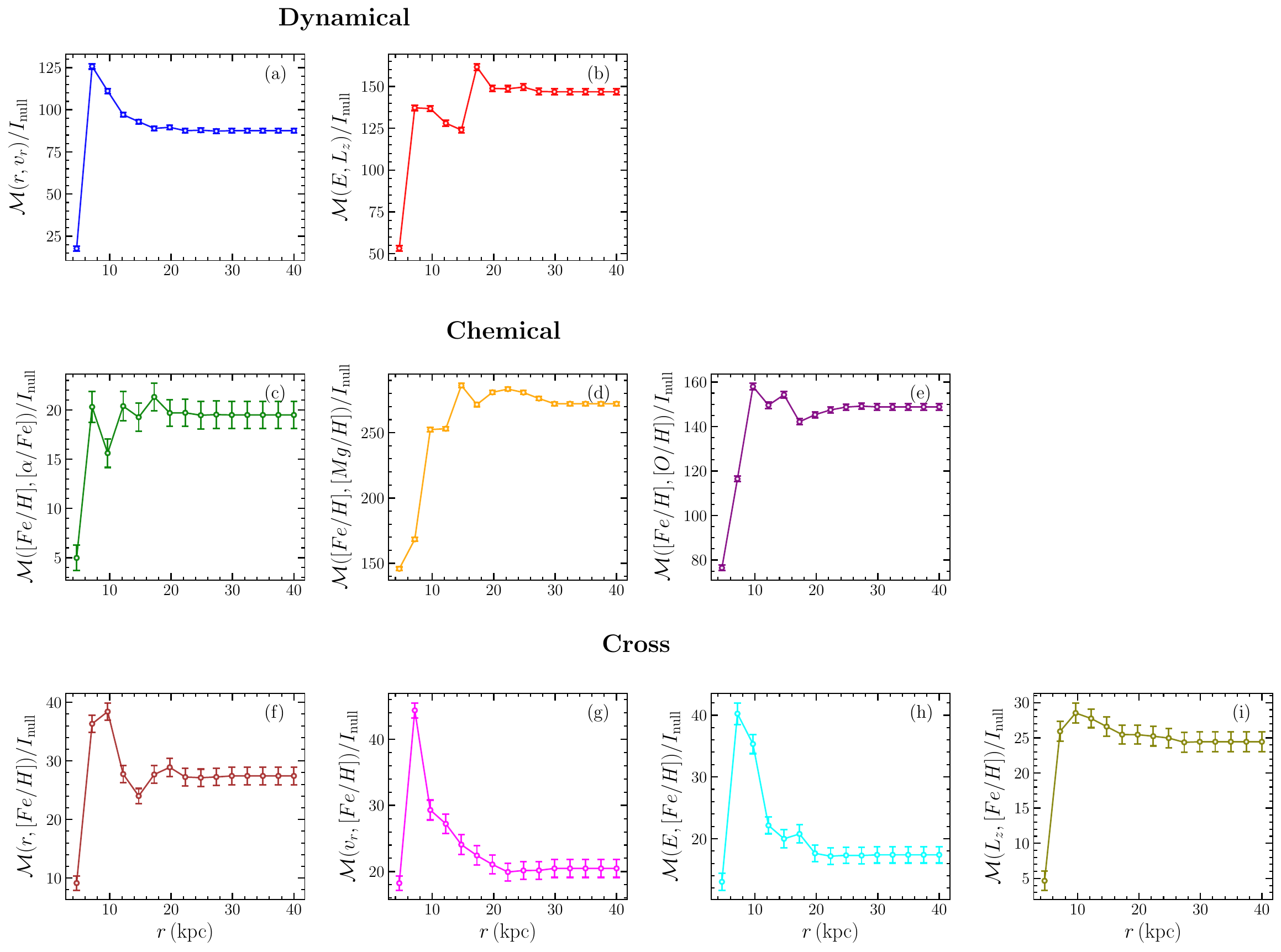}
\caption{Radial profiles of the cumulative stellar halo memory measured from APOGEE DR17 halo stars. Each profile is computed cumulatively using all stars interior to Galactocentric radius $r$, with error bars representing $1\sigma$ jackknife uncertainties. Despite exhibiting distinct radial evolution, the dynamical, chemical, and cross-memory observables all converge toward nearly constant values beyond $r\sim20$\,kpc, indicating the emergence of a statistically significant residual memory state in the outer stellar halo.}
\label{fig:significance}
\end{figure*}

\begin{figure}
\centering
\includegraphics[width=0.4\textwidth]{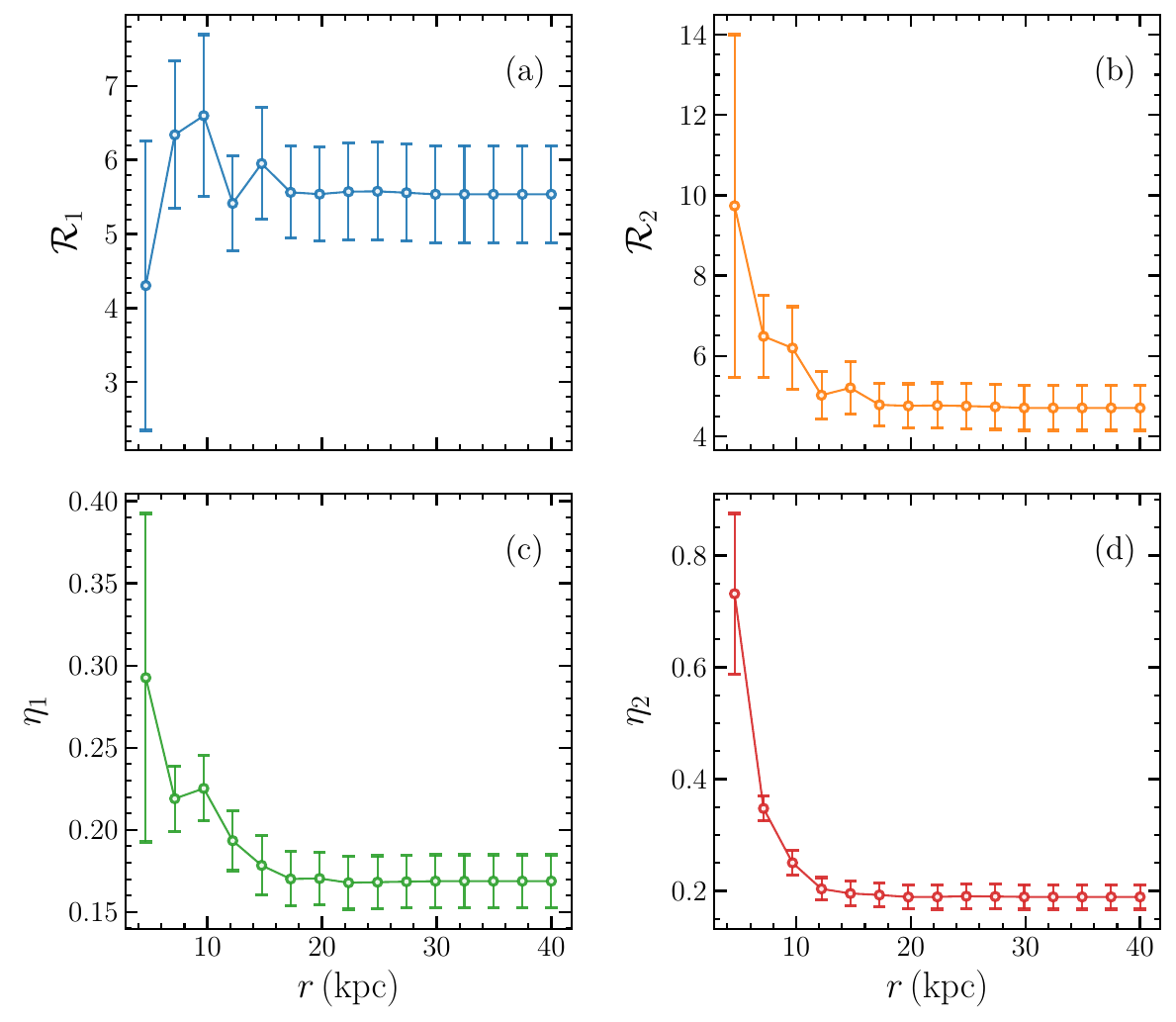}
\caption{Derived diagnostics describing the partitioning of stellar halo memory. Panels (a) and (b) show the memory dominance ratios, $\mathcal{R}_1=\mathcal{M}(r,v_r)/\mathcal{M}([\mathrm{Fe/H}],[\alpha/\mathrm{Fe}])$ and $\mathcal{R}_2=\mathcal{M}(E,L_z)/\mathcal{M}([\mathrm{Fe/H}],[\alpha/\mathrm{Fe}])$, respectively. Panels (c) and (d) show the corresponding coupling efficiencies, $\eta_1$ and $\eta_2$ (\autoref{eq:efficiency}). The surviving assembly memory is consistently dominated by dynamical correlations while retaining a finite chemo-dynamical coupling throughout the Galactic stellar halo.}
\label{fig:diagnostics}
\end{figure}

\subsection{Radial Evolution of Stellar Halo Memory}

\autoref{fig:memory} presents the radial profiles of the dynamical, chemical, and cross-memory observables constructed from the cumulative stellar sample. Although the individual observables exhibit markedly different radial behaviour, a common evolutionary pattern emerges. Most memory observables vary rapidly within the inner halo ($r\lesssim10$ kpc), where the cumulative stellar population changes most rapidly with increasing radius, before gradually approaching nearly constant residual memory state beyond $r\sim20$ kpc.

The dynamical memory observables, $\mathcal{M}(r,v_r)$ and $\mathcal{M}(E,L_z)$, both decrease with increasing radius after an initial peak, indicating that the strong phase-space coherence present in the inner halo becomes progressively diluted as additional stellar populations are incorporated into the cumulative sample. The decline is substantially steeper for $\mathcal{M}(E,L_z)$ than for $\mathcal{M}(r,v_r)$, suggesting that orbital correlations are more efficiently redistributed than radial phase-space correlations.

The chemical memory observables display a richer diversity of behaviour. While $\mathcal{M}([\mathrm{Fe/H}],[\mathrm{Mg/H}])$ and $\mathcal{M}([\mathrm{Fe/H}],[\mathrm{O/H}])$ decrease monotonically with radius, $\mathcal{M}([\mathrm{Fe/H}],[\alpha/\mathrm{Fe}])$ remains remarkably stable over most of the radial range, indicating that the coupling between global metallicity and $\alpha$-enhancement is largely preserved throughout the halo.

The cross-memory observables similarly evolve toward nearly stationary values despite exhibiting different radial trends in the inner halo. In particular, the strongest decline is observed for $\mathcal{M}(v_r,[\mathrm{Fe/H}])$, whereas $\mathcal{M}(L_z,[\mathrm{Fe/H}])$ evolves comparatively slowly. These differences demonstrate that the efficiency with which dynamical evolution redistributes chemo-dynamical information depends on the particular dynamical variable considered.

The most striking result, however, is not the diversity of the individual profiles, but their convergence. Despite tracing distinct physical processes, all dynamical, chemical, and cross-memory observables approach approximately constant values beyond $r\sim20$ kpc. This behaviour suggests that the outer stellar halo approaches a statistically stationary memory state in which a finite amount of assembly information survives despite extensive dynamical evolution.

\subsection{Statistical Significance of the Memory}

The significance of the measured memory is assessed by comparing every observable with its corresponding shuffled realization (\autoref{fig:significance}). Every dynamical, chemical, and cross-memory observable remains significantly above its null expectation, demonstrating that the surviving assembly memory is an intrinsic property of the stellar halo rather than an artifact of statistical estimation.

The normalized memory remains typically one to two orders of magnitude larger than the null expectation for the dynamical and cross-memory observables, while several chemical memory observables exceed the null level by factors approaching a few hundred. Such large excesses demonstrate that the measured memory cannot be attributed to finite-sample fluctuations or biases inherent to the mutual-information estimator. Instead, the observed correlations represent genuine statistical organization preserved within the stellar halo.

Importantly, the normalization by the null model preserves the qualitative radial behaviour identified in \autoref{fig:memory}. Observables that decrease in absolute memory continue to decrease after normalization, whereas those exhibiting nearly constant memory remain similarly stable. Consequently, the asymptotic behaviour observed beyond $r\sim20$ kpc does not arise because the measurements approach the estimator noise floor. Rather, it represents a physically significant residual memory that remains well above the statistical expectation for an uncorrelated stellar population.

\subsection{Partitioning the Residual Memory}

While \autoref{fig:memory} and \autoref{fig:significance} quantify the amount of surviving stellar halo memory, \autoref{fig:diagnostics} reveals how this memory is partitioned between its dynamical, chemical, and chemo-dynamical components.

The memory dominance ratios remain larger than unity over the entire radial range, indicating that the surviving memory is consistently dominated by dynamical rather than chemical correlations. Both $\mathcal{R}_1$ and $\mathcal{R}_2$ exhibit their largest values within the inner halo before declining and approaching nearly constant values at larger radii. The systematic difference between the two ratios further indicates that the relative importance of dynamical and chemical memory depends on the dynamical observable used to characterize the phase-space structure.

The coupling efficiencies display a qualitatively different evolution. Both $\eta_1$ and $\eta_2$ decrease rapidly over the inner $\sim10$--15 kpc before reaching nearly constant values beyond $\sim20$ kpc. This behaviour indicates that the coupling between stellar chemistry and dynamics weakens as progressively larger regions of the halo are included, consistent with the cumulative effects of phase mixing and hierarchical assembly. Nevertheless, the coupling efficiency remains significantly above zero at all radii, demonstrating that the dynamical and chemical memories never become completely independent.

Taken together, Figure~1--3 reveal a coherent picture of the surviving memory of the Galactic stellar halo. Individual memory observables exhibit diverse radial evolution reflecting the different physical processes governing orbital dynamics, chemical enrichment, and their mutual coupling. Nevertheless, all memory observables converge toward a common statistically significant residual state beyond approximately $20$ kpc, which we refer to here as the \emph{residual memory state} (RMS) of the stellar halo. This convergence does not indicate the disappearance of assembly information. Rather, it marks the emergence of a nearly stationary statistical configuration in which a finite amount of dynamical, chemical, and chemo-dynamical memory continues to survive. The residual memory remains dominated by dynamical correlations while preserving a measurable chemo-dynamical coupling, suggesting that phase mixing redistributes, rather than completely erases, the information imprinted during galaxy assembly. The outer stellar halo therefore appears not as an information-free remnant of dynamical evolution, but as a statistically stationary repository of the surviving memory of the Milky Way's assembly history.

\section{Conclusions}

We have introduced a unified information-theoretic framework for quantifying the memory preserved in the Galactic stellar halo using observational data. By treating the stellar halo as an information reservoir, our approach provides a common statistical language for describing how galaxy assembly is encoded in stellar dynamics, chemical abundances, and their mutual coupling. Rather than analysing these observables independently, the present framework measures their shared information content and interprets it as distinct, yet complementary, memory reservoirs of galaxy formation.

Applying this framework to APOGEE DR17 halo stars, we obtain the following principal results:

\begin{itemize}
\item We introduce stellar halo memory as a new observable of Galactic archaeology and demonstrate that it can be robustly quantified using the excess mutual information between stellar dynamical and chemical observables.

\item We identify three complementary classes of memory i.e. dynamical memory, chemical memory, and cross memory, which provide a physically motivated decomposition of the information preserved within the stellar halo.

\item All memory observables exhibit distinct radial evolution in the inner halo but converge toward a common statistically significant residual memory state beyond approximately $20$~kpc, suggesting that the outer halo retains a finite memory of its assembly history despite extensive dynamical evolution.

\item Comparison with randomized realizations demonstrates that all measured memory observables exceed their corresponding null expectations by large factors, confirming that the detected memory represents genuine astrophysical structure rather than statistical fluctuations or estimator bias.

\item The memory dominance ratios indicate that the surviving assembly memory is consistently dominated by dynamical correlations, whereas the coupling efficiencies reveal that a finite chemo-dynamical association persists throughout the halo, even after substantial phase mixing.
\end{itemize}

Overall, these results suggest that the present-day stellar halo is not merely a dynamically mixed remnant of hierarchical assembly, but a statistically organized system that preserves measurable information about its formation history. The emergence of a common residual memory state across all dynamical, chemical, and cross-memory observables further suggests that phase mixing redistributes, rather than completely erases, the information imprinted during galaxy formation.

The framework presented here naturally extends beyond the Milky Way. A particularly important next step is to apply the same formalism to cosmological hydrodynamical simulations such as IllustrisTNG \citep{volker18}, AURIGA \citep{monachesi19}, and FIRE-2 \citep{wetzel23}, where the complete formation histories of individual galaxies are known. Such analyses will allow the evolution of stellar halo memory to be followed across cosmic time and will establish how the different memory reservoirs depend on halo mass, concentration, spin, ex-situ fraction, merger history, environment, and redshift. These simulations will also provide the opportunity to investigate whether the residual memory state identified here represents a generic outcome of hierarchical galaxy formation or a distinctive property of the Milky Way.

Finally, this work establishes stellar halo memory as a new observable for Galactic archaeology, providing a unified statistical framework for quantifying how the assembly memory imprinted during galaxy formation is preserved, redistributed, and ultimately remembered by the stellar halo.

\section{Acknowledgments}
Funding for the Sloan Digital Sky Survey IV has been provided by the Alfred P. Sloan Foundation, the U.S. Department of Energy Office of Science, and the Participating Institutions. 

SDSS-IV acknowledges support and resources from the Center for High-Performance Computing at the University of Utah. The SDSS web site is \url{www.sdss.org}.

SDSS-IV is managed by the Astrophysical Research Consortium for the Participating Institutions of the SDSS Collaboration including the Brazilian Participation Group, the Carnegie Institution for Science, Carnegie Mellon University, the Chilean Participation Group, the French Participation Group, the Harvard-Smithsonian Center for Astrophysics, the Instituto de Astrof\'isica de Canarias, the Johns Hopkins University, the Kavli Institute for the Physics and Mathematics of the Universe (IPMU) / University of Tokyo, the Lawrence Berkeley National Laboratory, the Leibniz Institute for Astrophysics Potsdam (AIP), the Max-Planck-Institut f\"ur Astronomie (MPIA Heidelberg), the Max-Planck-Institut f\"ur Astrophysik (MPA Garching), the Max-Planck-Institut f\"ur Extraterrestrische Physik (MPE), National Central University, National Astronomical Observatory of China (NAOC), National Tsing Hua University, the Ohio State University, the Pennsylvania State University, the Shanghai Astronomical Observatory, the United Kingdom Participation Group, the Universidad Nacional Aut\'onoma de M\'exico, the University of Arizona, the University of Colorado Boulder, the University of Edinburgh, the University of Florida, the University of Michigan, the University of Notre Dame, the University of Oviedo, the University of Tokyo, the University of Washington, the University of Wisconsin, the University of Wyoming, and Yale University.

BP acknowledges the financial support from the Anusandhan National Research Foundation, Government of India through the project ANRF/ARG/2025/000535/PS. BP also acknowledges IUCAA, Pune for providing support through associateship programme. AM acknowledges the University Grants Commission (UGC), Government of India, for support through a Junior Research Fellowship.

\section{Data availability}
The APOGEE DR17 data used in this study are publicly available at \url{https://www.sdss4.org/dr17/}. The derived data products generated in this work are available from the authors upon reasonable request.

\bibliographystyle{aasjournal}
\bibliography{refs.bib}{}

\end{document}